\documentclass[aps,prd,twocolumn,nofootinbib,superscriptaddress,floatfix]{revtex4-2}

\usepackage{amsmath,amssymb,bm}
\usepackage{graphicx}
\usepackage{booktabs}
\usepackage{siunitx}
\usepackage{hyperref}
\usepackage{xcolor}
\usepackage{microtype}

\hypersetup{colorlinks=true,citecolor=blue,linkcolor=blue,urlcolor=blue}
\newcommand{\jhat}{\widehat{j}}

\newcommand{\KR}{Kalb--Ramond}
\newcommand{\dd}{\mathrm{d}}
\usepackage{orcidlink}
\begin{document}

\title{Slow rotation of a localized Kalb--Ramond wormhole: Wald charges and fractional quadrupolar hair}

\author{Sardor~Murodov\orcidlink{0000-0003-2360-4475}}
\email{s.murodov@newuu.uz}
\affiliation{New Uzbekistan University, Movarounnahr Street 1, Tashkent 100000, Uzbekistan}
\affiliation{Institute of Fundamental and Applied Research, National Research University TIIAME, Kori Niyoziy 39, Tashkent 100000, Uzbekistan}
\date{\today}

\begin{abstract}
We construct an action-consistent slow-rotation branch, through second order, for a traversable wormhole supported by a phantom scalar, a scalar-localized radial string sector, and a Kalb--Ramond two-form coupled nonminimally to both the Ricci scalar and Ricci tensor. For the regular benchmark branch, covariant asymptotic analysis shows a vanishing ADM mass, while the first-order axial mode carries finite Iyer--Wald angular momentum. The nonminimal contributions to the angular charge decay at infinity, so the covariant and metric definitions of angular momentum agree. At second order, no nonzero mass correction is resolved; at fixed asymptotic scalar charge, rotation increases the throat area and makes the throat oblate. The most distinctive result arises in the quadrupolar sector. A long-range polar Kalb--Ramond response couples to the localized background and generates a matter-supported metric tail that decays more slowly than the ordinary vacuum quadrupole. The same noninteger hierarchy appears in curvature and reduces the differentiability of the conformal completion at spatial infinity. Consequently, the Geroch--Hansen and asymptotically Cartesian mass-centered quadrupole constructions are not applicable in their standard smooth form to this branch, even though its global mass and angular-momentum charges remain well defined. This separates charge-level asymptotic flatness from smooth multipolar asymptotics and identifies fractional Kalb--Ramond hair as the leading rotational quadrupolar signature of the benchmark solution.
\end{abstract}

\maketitle

\section{Introduction}
\label{sec:intro}
Traversable Lorentzian wormholes provide a direct setting in which spacetime topology, matter support, global charges, and asymptotic structure must be treated together.  The Ellis--Bronnikov solutions supplied the simplest scalar-supported examples \cite{Ellis1973,Bronnikov1973}, while the Morris--Thorne construction made explicit the throat and flare-out conditions required for traversability \cite{MorrisThorne1988}.  In four-dimensional general relativity these conditions are tied to violations of the null energy condition, as emphasized by thin-shell constructions, local throat theorems, and topological-censorship results \cite{MorrisThorneYurtsever1988,Visser1989,FriedmanSchleichWitt1993,HochbergVisser1997,HochbergVisser1998,LemosLoboOliveira2003}.  A physically meaningful rotating wormhole should therefore be obtained from a specified matter--gravity system rather than from a metric ansatz alone.

Stability provides a second motivation for studying rotation.  The massless ghost-scalar wormhole is dynamically delicate: nonlinear evolutions can lead to collapse or expansion, and linear analyses identify an unstable radial mode for broad classes of static solutions \cite{ShinkaiHayward2002,GonzalezGuzmanSarbach2009,ToriiShinkai2013}.  The standard perturbative framework for rotation is the Hartle--Thorne expansion \cite{Hartle1967,HartleThorne1968}.  Teo introduced a stationary axisymmetric wormhole geometry \cite{Teo1998}, while self-consistent phantom-scalar wormholes were subsequently constructed through first and second order in slow rotation \cite{SushkovKashargin2008,KasharginSushkov2008}; related regular rotational modes were analyzed in Ref.~\cite{AzregAinou2012}.  Fully nonlinear rotating Ellis wormholes and rotating matter-supported extensions have since shown that spin can modify throat geometry, global charges, ergoregions, and the stability balance \cite{DzhunushalievEtAl2013,KleihausKunz2014,ChewKleihausKunz2016,ChewKleihausKunz2018STT,ChewEtAl2019,HoffmannEtAl2018PLB,HoffmannEtAl2018PRD,BlazquezSalcedoChewKunz2018,TsukamotoKokubu2018,AzadEtAl2024}.  Recent analytic Teo-type constructions further reinforce that rotation by itself is not a sufficient novelty criterion; what matters is the response predicted by a specified covariant action \cite{BaticDutykhSukaiti2026CQG,BaticDutykhSukaiti2026EPJC,Errehymy2026}.

Modified-gravity interactions provide additional mechanisms for sustaining or reshaping wormhole throats.  Higher-curvature couplings can shift part of the effective energy-condition violation into the gravitational sector \cite{KantiKleihausKunz2011,KantiKleihausKunz2012,HarkoLoboMakSushkov2013,LoboOliveira2009}, while form fields and topological defects supply further nontrivial sources \cite{BarrosLobo2018,Jusufi2018}.  In particular, the Einstein--scalar--Gauss--Bonnet and related wormholes studied by Ibadov, Kleihaus, Kunz, and Murodov exhibit nontrivial domains of existence, throat structures, and global properties \cite{IbadovKleihausKunzMurodov2020,IbadovKleihausKunzMurodov2021,IbadovKleihausKunzMurodov2022}.  The construction developed below is complementary: the additional gravitational degree of freedom is an antisymmetric two-form with two independent nonminimal curvature couplings, localized together with a radial string sector by a scalar-dependent profile.

The Kalb--Ramond (KR) two-form was introduced as the antisymmetric tensor field that couples naturally to strings \cite{KalbRamond1974}.  A nonzero vacuum expectation value of an antisymmetric tensor can generate spontaneous Lorentz breaking in the presence of nonminimal curvature interactions \cite{AltschulBaileyKostelecky2010}.  In gravitational applications, background KR fields have produced modified black holes and traversable wormholes \cite{LessaSilvaMalufAlmeida2020,LessaOliveiraSilvaAlmeida2021,MalufMuniz2022}, and static or slowly rotating black-hole sectors have been developed in several formulations, including non-commutative and matter-coupled extensions \cite{Yang2023,LiuWuWang2024,LiuWuWang2025,LiuWuWei2025,AraujoFilho2025NonCommutativeKR,AhmedSilva2026DyonicModMax}.  Their particle dynamics, neutrino phenomenology, optical and lensing signatures, thermodynamic behavior, evaporation properties, and observational constraints have also been investigated \cite{AtamurotovEtAl2022,AraujoFilho2025,AraujoFilho2025ParticleMotionKR,Shi2025NeutrinoKR,AraujoFilho2025AntisymmetricTensorLensing,Pereira2026ChargedKRLensing,AlBadawiAhmedSakalli2025KRModMax,AhmedAlBadawiSakalli2026KRConstraints}.  Recent work has emphasized two subtleties directly relevant here: the asymptotic behavior of KR/torsional hair can depend sensitively on the accompanying string dynamics \cite{GarciaAndrade2026}, and nonminimal curvature couplings can complicate the identification of physical gravitational charges from metric parameters alone \cite{YuLyuHuheLi2026}.  Our radial string component is motivated by the covariant string-cloud/string-fluid framework \cite{Letelier1979,SmalleyKrisch1997}, with a scalar-dependent tension chosen to localize the string and KR sectors simultaneously.

These observations make the asymptotic analysis as important as the local throat geometry.  In a diffeomorphism-invariant theory with nonminimal curvature couplings, mass and angular momentum should be checked with covariant phase-space/Noether-charge methods rather than inferred only from metric coefficients \cite{LeeWald1990,Wald1993,IyerWald1994}.  Likewise, the usual Geroch--Hansen and Thorne multipole hierarchies require a sufficiently regular asymptotic expansion and conformal completion \cite{Geroch1970,Hansen1974,Thorne1980,Gursel1983,BeigSimon1981,FodorHoenselaersPerjes1989,BackdahlHerberthson2005}.  Modern Noether-charge and nonvacuum analyses make these regularity assumptions explicit and show why finite global charges need not guarantee a standard smooth higher-multipole expansion \cite{CompereOliveriSeraj2018,CanoGanchevMayerson2022,Mayerson2023Multipoles,BongaGrantPrabhu2020}.

Against this background, we construct an action-consistent slow-rotation branch through $O(J^2)$ for a traversable wormhole supported by a phantom scalar, a localized radial Nambu--Goto/string-cloud sector, and a KR two-form coupled nonminimally to both $R$ and $R_{\mu\nu}$.  The fixed-norm constraint, axial two-form response, second-order string bending, and curvature contributions to the Wald charges are treated within the same variational system.  For the regular benchmark branch, the global mass and angular momentum charges remain well defined, whereas the leading rotational quadrupolar response develops a noninteger far-field decay that precedes the ordinary vacuum quadrupole.  This fractional tail also appears in curvature and reduces the differentiability of the conformal completion, so the standard smooth Geroch--Hansen/ACMC quadrupole construction is not applicable to this branch.  To our knowledge, this combination of two independent KR curvature couplings, scalar localization, action-derived second-order matter response, covariant charge normalization, and an explicit asymptotic-smoothness test has not been treated in the existing KR or rotating-wormhole constructions cited above.

The paper is organized as follows.  Section~\ref{sec:action} introduces the covariant action and localization mechanism, and Sec.~\ref{sec:static} constructs the static wormhole background.  Section~\ref{sec:firstorder} develops the regular first-order rotational sector.  The second-order monopole and constrained quadrupolar systems are analyzed in Secs.~\ref{sec:monopole} and \ref{sec:l2}, respectively, followed by the fractional KR quadrupolar tail in Sec.~\ref{sec:fractional}.  Section~\ref{sec:multipoles} examines the asymptotic charges and the obstruction to a standard smooth multipole hierarchy, while Sec.~\ref{sec:validity} states the perturbative domain of validity and limitations.  We discuss the physical interpretation and comparisons with earlier work in Sec.~\ref{sec:discussion} and summarize the main conclusions in Sec.~\ref{sec:conclusions}.  Technical formulae are collected in the Appendix.

We use signature $(-,+,+,+)$ and geometrized units $G=c=1$.  The throat radius $r_0$ sets the length scale; dimensionless radial variables are understood unless $r_0$ is restored explicitly.

\section{Action and localization mechanism}
\label{sec:action}
We use a single diffeomorphism-invariant action for the static and rotating sectors.  In gravitational units the bulk part is
\begin{equation}
S_{\rm bulk}=\frac{1}{2\kappa}\int \dd^4x\sqrt{-g}\,\mathcal L_{\rm bulk},
\label{eq:full_bulk_action}
\end{equation}
with
\begin{align}
\mathcal L_{\rm bulk}={}&R
-\frac1{12}H_{\alpha\beta\gamma}H^{\alpha\beta\gamma}
+\frac12\nabla_\mu\phi\nabla^\mu\phi\nonumber\\
&-\Lambda_B\big(B^2+b_0^2F(\phi)\big)
+\xi_1B^2R+\xi_2X^{\mu\nu}R_{\mu\nu}.
\label{eq:Lbulk}
\end{align}
where
\begin{align}
H&=\dd B, & B^2&=B_{\mu\nu}B^{\mu\nu},\\
X^{\mu\nu}&=B^{\mu\lambda}B^\nu{}_{\lambda}.&&
\end{align}
The positive sign of the scalar kinetic term in Eq.~\eqref{eq:full_bulk_action}, with signature $(-,+,+,+)$, makes $\phi$ a phantom scalar.  The field $\Lambda_B$ is a Lagrange multiplier enforcing
\begin{equation}
B^2=-b_0^2F(\phi),
\qquad
F(\phi)=\cos^2\!\left(\frac{\phi}{f_{\rm loc}}\right).
\label{eq:constraint_potential}
\end{equation}

The radial string sector is represented by a continuous Nambu--Goto congruence,
\begin{equation}
S_{\rm sc}=-\frac{1}{2\kappa}\int \dd^2y\,\nu(y)
\int \dd^2\sigma\,\mu_{\rm eff}(\phi)
\sqrt{-\det\gamma_{AB}},
\label{eq:NG_action}
\end{equation}
where $y$ labels the worldsheets, $\nu(y)$ is their conserved label density,
$\gamma_{AB}=g_{\mu\nu}\partial_A X^\mu\partial_B X^\nu$ is the induced metric, and the normalization of $\nu$ is absorbed into $\mu_{\rm eff}$.  In the static radial congruence this gives
\begin{equation}
T_{\hat a\hat b}^{\rm sc}
=\mathrm{diag}(\sigma,-\sigma,0,0),
\qquad
\sigma=\frac{\mu_{\rm eff}(\phi)}{R^2}.
\label{eq:stringstress}
\end{equation}
For the benchmark branch we choose
\begin{equation}
\mu_{\rm eff}(\phi)
=\left[\frac14+\frac38\cos^2\left(\frac{\phi}{\sqrt3}\right)\right]F(\phi).
\label{eq:mueff}
\end{equation}
The bracketed prefactor in Eq.~\eqref{eq:mueff} is smooth and positive, while the common factor $F$ forces both the string tension and the two-form condensate to switch off asymptotically.  This choice is therefore a simple benchmark for a co-localized throat-supported matter sector rather than a claimed fundamental form.  Equation~\eqref{eq:mueff} is part of the benchmark model definition; we do not claim universality with respect to other localization profiles.

Varying Eq.~\eqref{eq:full_bulk_action} with respect to the two-form gives
\begin{align}
0={}&\nabla_\lambda H^\lambda{}_{\mu\nu}
+\xi_2 B_\mu{}^\rho R_{\nu\rho}
-\xi_2 B_\nu{}^\rho R_{\mu\rho}\nonumber\\
&+2\xi_1 R B_{\mu\nu}
-4\Lambda_B B_{\mu\nu}.
\label{eq:KR_eom}
\end{align}
The factor of two in the $\xi_1$ current follows from
$\delta(B^2)=2B^{\mu\nu}\delta B_{\mu\nu}$ and is essential for the stationary branch.  On the constraint surface the multiplier contributes
\begin{equation}
T^{\rm pot}_{\mu\nu}=4\Lambda_B B_{\mu\alpha}B_\nu{}^\alpha.
\label{eq:Tpot}
\end{equation}
The phantom-scalar stress tensor is
\begin{equation}
T^\phi_{\mu\nu}
=-\left(\nabla_\mu\phi\nabla_\nu\phi
-\frac12g_{\mu\nu}(\nabla\phi)^2\right).
\label{eq:Tphi}
\end{equation}

It is convenient to define
\begin{equation}
\alpha\equiv\xi_2b_0^2,
\qquad
\gamma\equiv\frac{\xi_1}{\xi_2}.
\label{eq:alphagamma}
\end{equation}
Throughout the numerical benchmark we fix $\alpha=0.5$ and $r_0=1$.  All matter, metric, fixed-norm and bending perturbations used below are obtained by expanding Eqs.~\eqref{eq:full_bulk_action} and \eqref{eq:NG_action}; no independent rotating stress tensor is imposed by hand.

\section{Static localized wormhole}
\label{sec:static}
The static geometry is written in proper-radial gauge,
\begin{equation}
\dd s^2=-N(l)^2\dd t^2+\dd l^2+R(l)^2\dd\Omega^2,
\label{eq:staticmetric}
\end{equation}
with the electric two-form
\begin{equation}
B^{(0)}=-\beta(l)\,\dd t\wedge\dd l.
\label{eq:staticB}
\end{equation}
The fixed-norm condition gives
\begin{equation}
\beta=N b_0\sqrt{\frac{F}{2}}.
\label{eq:beta_constraint}
\end{equation}
At the reflection-symmetric throat we impose
\begin{equation}
R(0)=r_0,
\qquad R'(0)=N'(0)=0,
\qquad \phi(0)=0,
\label{eq:throatstatic}
\end{equation}
while asymptotic localization requires $F\to0$ and $\phi\to\pi f_{\rm loc}/2$ on each asymptotic end.

For each trial $\gamma$ the static equations are solved first, with the multiplier obtained from the static $\mathrm{KR}_{01}$ equation.  The remaining stationary root is then fixed by requiring that the two-dimensional decaying axial stable subspace satisfy the independent throat regularity conditions.  The two resulting residuals determine $(\gamma,\xi_2)$.  Thus the coupling pair quoted below is selected by the \emph{stationary} boundary-value problem, not by the static equations alone.  Covariant static residuals and the scalar Noether identity are reported below as numerical consistency checks.

The regular branch found in the scanned neighborhood is summarized in Table~\ref{tab:parameters}.  We emphasize that this is a regular codimension-two root in the explored $(\gamma,\xi_2)$ domain; global uniqueness over the entire coupling plane is not claimed.  On the final high-accuracy static background, representative covariant residuals are $|E_{00}|\simeq3.8\times10^{-16}$, $|E_{11}|\simeq5.9\times10^{-12}$, and $|E_{22}|\simeq5.3\times10^{-16}$, while the scalar Noether identity is satisfied at the $10^{-10}$ level.  These numbers are quoted only as numerical consistency checks.

\begin{table}[t]
\caption{Benchmark parameters of the final production branch.  All quoted numerical results in this paper use this refined root.}
\label{tab:parameters}
\begin{ruledtabular}
\begin{tabular}{lc}
quantity & value\\
\hline
$\alpha=\xi_2b_0^2$ & $0.5$\\
$\gamma=\xi_1/\xi_2$ & $0.3187655084985281$\\
$\xi_2$ & $10.658802724675182$\\
$\xi_1$ & $3.3976586705165808$\\
$f_{\rm loc}/r_0$ & $1.7474165128837902$\\
$p_+$ & $0.7936049731094987$\\
\end{tabular}
\end{ruledtabular}
\end{table}

\begin{figure}[t]
\includegraphics[width=\columnwidth]{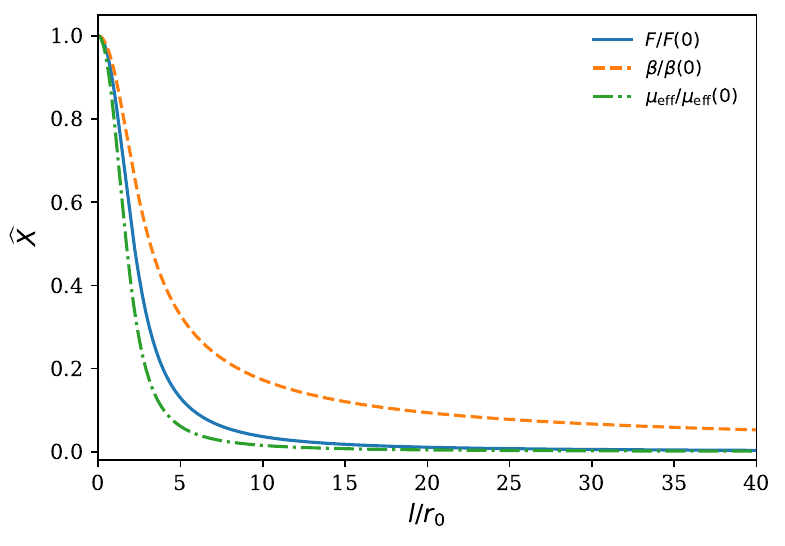}
\caption{Localization of the static matter sector.  The ordinate is $\widehat X\equiv X(l)/X(0)$; the three curves correspond to $X=F$, $\beta$, and $\mu_{\rm eff}$, respectively.}
\label{fig:localization}
\end{figure}

Figure~\ref{fig:localization} shows the localization function, two-form amplitude, and effective string tension.  The outer scalar mode behaves as
\begin{equation}
F\sim R^{-2p_+},
\qquad
\beta\sim R^{-p_+},
\qquad p_+>\frac12.
\label{eq:staticfalls}
\end{equation}

\begin{figure}[t]
\includegraphics[width=\columnwidth]{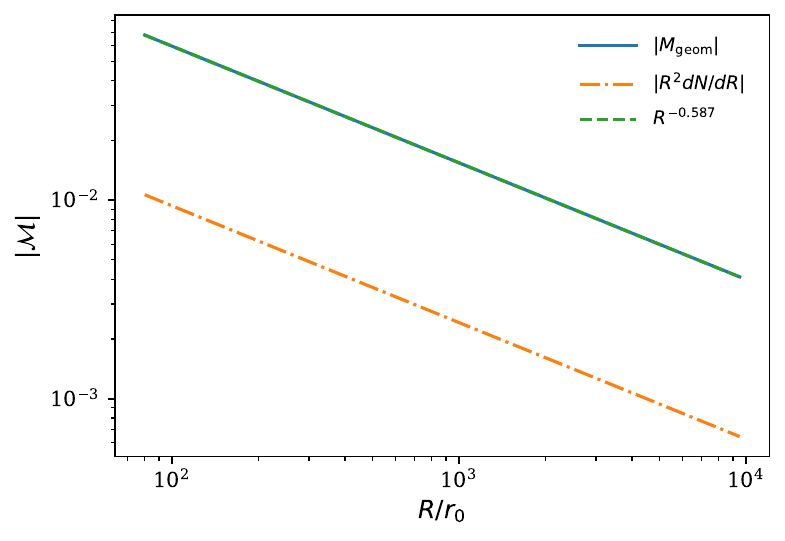}
\caption{Independent asymptotic mass estimators.  Here $\mathcal M$ denotes either $M_{\rm geom}=R(1-R'^2)/2$ or $M_N=R^2N'/R'$.  Both $|\mathcal M|$ decay with the predicted localization exponent $R^{1-2p_+}$ and tend to zero.}
\label{fig:massless}
\end{figure}

A useful check is the asymptotic mass.  In proper-radial gauge a geometric mass estimator is
\begin{equation}
M_{\rm geom}(R)=\frac{R}{2}\left(1-R'^2\right).
\label{eq:Mgeom}
\end{equation}
The independent lapse estimator is $R^2\dd N/\dd R$.  Both fall with the same noninteger power,
\begin{align}
M_{\rm geom},\ R^2\frac{\dd N}{\dd R}
&\propto R^{1-2p_+},\nonumber\\
1-2p_+&=-0.5872099462\ldots .
\label{eq:massfall}
\end{align}
with no finite $1/R$ mass coefficient.  Figure~\ref{fig:massless} displays this decay.  We therefore obtain
\begin{equation}
M_{\rm ADM}^{(0)}=0.
\label{eq:M0zero}
\end{equation}
This result will make the throat scale, rather than $M$, the natural normalization for angular momentum.

\section{First-order rotation and the regular axial branch}
\label{sec:firstorder}
We introduce slow rotation with a bookkeeping parameter $\epsilon$.  At first order the coframe can be written
\begin{align}
e^{\hat0}&=N\dd t,
& e^{\hat1}&=\dd l,
& e^{\hat2}&=R\dd\theta,\\
e^{\hat3}&=R\sin\theta\,\dd\phi
-\epsilon\frac{h(l)}{R}\sin\theta\,\dd t,
\end{align}
so that
\begin{equation}
g_{t\phi}=-\epsilon h(l)\sin^2\theta.
\label{eq:gtphi}
\end{equation}
The axial \KR{} perturbation is parameterized by
\begin{align}
B_{l\phi}^{(1)}&=\beta V\sin^2\theta,
&
B_{\theta\phi}^{(1)}&=\beta w\sin\theta\cos\theta.
\label{eq:axialB}
\end{align}
For regular numerical evolution we use
\begin{equation}
\begin{aligned}
K&=\xi_2\beta^2V,
&\bar U&=\sqrt{\xi_2}\beta w,\\
\bar Q&=\sqrt{\xi_2}\left[(\beta w)'-2\beta V\right].
\end{aligned}
\label{eq:regularvars}
\end{equation}
with state $X=(h,h',K,K',\bar U,\bar Q)$.  Reflection parity and regularity are imposed exactly at the throat.  We normalize the internal eigenmode by $\bar U(0)=1$; the resulting derivatives are
\begin{equation}
h'(0)=-3.7796294322,
\qquad
K'(0)=3.1200804972.
\label{eq:axialthroat}
\end{equation}
The acceptable far modes have exponents approximately $-1$ and $-2.7087$.  To avoid contamination by growing modes, the asymptotic stable subspace is initialized in the far zone and transported inward.  Direct substitution back into the untransformed $t\phi$ Einstein equation and the two axial \KR{} equations closes at floating-point/interpolation accuracy.

\begin{figure}[t]
\includegraphics[width=\columnwidth]{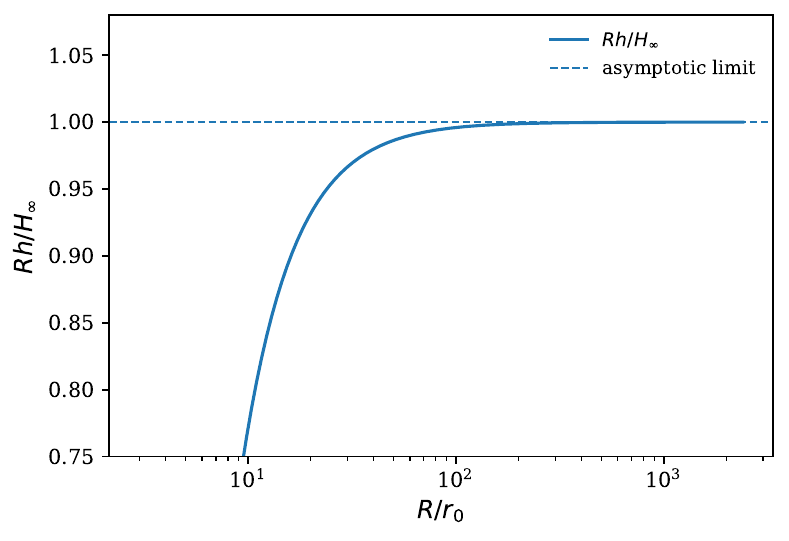}
\caption{First-order frame dragging.  The combination $Rh$ approaches the finite asymptotic coefficient $H_\infty$.}
\label{fig:dragging}
\end{figure}

The frame-dragging mode approaches
\begin{equation}
Rh\longrightarrow H_\infty^{(t)}=0.6846925027.
\label{eq:Hcoord}
\end{equation}
The small finite-domain lapse normalization $N_\infty=0.9999906327$ is removed by $t_\infty=N_\infty t$, giving
\begin{equation}
H_\infty=0.6846989165.
\label{eq:Hphys}
\end{equation}
The convergence is shown in Fig.~\ref{fig:dragging}.

\subsection{Iyer--Wald angular momentum}
\begin{figure}[t]
\includegraphics[width=\columnwidth]{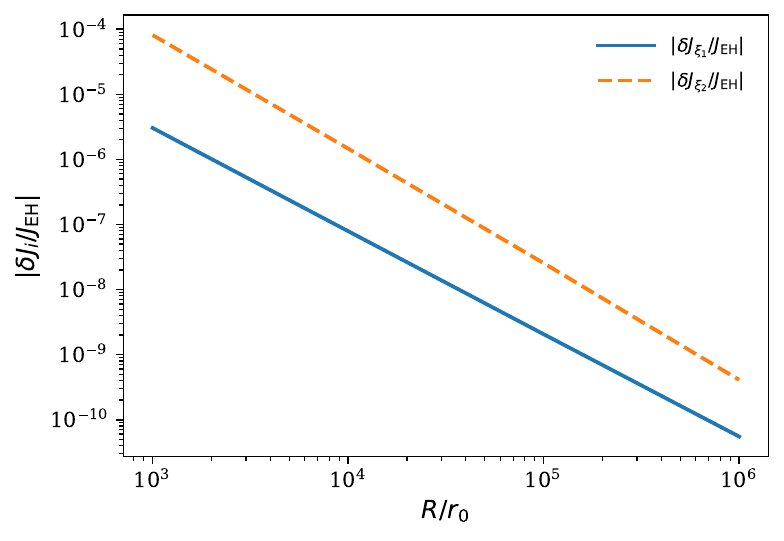}
\caption{Decay of the nonminimal axial Iyer--Wald contributions relative to the Einstein term.  The solid and dashed curves show $|\delta J_{\xi_1}/J_{\rm EH}|$ and $|\delta J_{\xi_2}/J_{\rm EH}|$, respectively; the distinct line styles keep the comparison unambiguous in grayscale.}
\label{fig:wald}
\end{figure}

Because of the curvature couplings in Eq.~\eqref{eq:full_bulk_action}, the $1/R$ coefficient in $g_{t\phi}$ must be checked against the full covariant surface charge.  For the scalar Lagrangian define $P^{abcd}=\partial\mathcal L/\partial R_{abcd}$.  In the present model
\begin{align}
P^{abcd}={}&\frac{1+\xi_1B^2}{4\kappa}
\left(g^{ac}g^{bd}-g^{ad}g^{bc}\right)\nonumber\\
&+\frac{\xi_2}{8\kappa}\left(
 g^{ac}X^{bd}-g^{ad}X^{bc}
-g^{bc}X^{ad}+g^{bd}X^{ac}\right).
\label{eq:Pabcd}
\end{align}
The Iyer--Wald potential contains both the $P\nabla\varphi$ and $\varphi\nabla P$ pieces; both are included in the expressions below.  For the axial Killing field $\varphi^a=(\partial_\phi)^a$, the three curvature-sector radial integrands, after dropping a common normalization, reduce at $O(\epsilon)$ to
\begin{align}
{\cal I}_{\rm EH}
&=\frac{-Rh'+2hR'}{4N^2R}\sin^2\theta,\\
\frac{{\cal I}_{\xi_1}}{{\cal I}_{\rm EH}}
&=-\frac{2\xi_1\beta^2}{N^2}
=\xi_1B^2=-\alpha\gamma F,\\
\frac{{\cal I}_{\xi_2}}{{\cal I}_{\rm EH}}
&=\frac{RK'}{-Rh'+2hR'}.
\label{eq:wald_ratios}
\end{align}
Localization gives $F\sim R^{-2p_+}$ and $K\sim R^{-(2+p_+)}$, so both ratios vanish.  Numerically their combined magnitude is approximately $4.6\times10^{-10}$ of the Einstein contribution by $R\sim10^6$ (Fig.~\ref{fig:wald}).  Since $H^{(0)}=0$, the kinetic two-form sector also leaves no independent linear surface charge under the present boundary conditions.  Thus
\begin{equation}
J_{\rm Wald}=J_{\rm metric}.
\label{eq:Jwald}
\end{equation}
Matching Eq.~\eqref{eq:gtphi} to $g_{t_\infty\phi}=-2J\sin^2\theta/R+\cdots$ yields
\begin{equation}
\jhat\equiv\frac{J}{r_0^2}
=0.3423494582\,\epsilon+O(\epsilon^3),
\label{eq:jhat}
\end{equation}
so that
\begin{equation}
\epsilon=2.920991916\,\jhat+O(\jhat^3).
\label{eq:epsj}
\end{equation}
Because $M_{\rm ADM}=0$, the Kerr-like parameter $J/M^2$ is undefined; $\jhat$ is the appropriate dimensionless spin for this throat-scale branch.

\section{Second-order monopole backreaction}
\label{sec:monopole}
At $O(\epsilon^2)$ the monopole deformations are introduced in signed proper-radial gauge as
\begin{equation}
\begin{aligned}
N&\to N(1+\epsilon^2 n_0),
& R&\to R(1+\epsilon^2 k_0),\\
\phi&\to\phi+\epsilon^2\phi_0.
\end{aligned}
\label{eq:monopole_ansatz}
\end{equation}
The quadratic sources include the metric frame-dragging contribution, the $H^2$ stress generated by the axial two-form, the fixed-norm electric correction, both nonminimal curvature sectors, the scalar response, and the perturbed Nambu--Goto density.  We define the asymptotic scalar-tail charge by
\begin{equation}
\phi(R)=\phi_\infty-Q_\phi R^{-p_+}+o(R^{-p_+}),
\qquad \phi_\infty=\frac{\pi f_{\rm loc}}{2},
\label{eq:Qphi}
\end{equation}
and remove the static scale zero mode so that $Q_\phi$ is held fixed on the rotating branch.

Under domain enlargement the shift of the coupling ratio is consistent with zero within the numerical convergence level,
\begin{equation}
\delta\gamma^{(2)}\simeq0.
\label{eq:dgamma}
\end{equation}
A gauge-invariant geometric mass estimator for the monopole correction is
\begin{equation}
\delta M_{\rm geom}
=Rk_0\frac{1-R'^2}{2}
-RR'\left(R'k_0+Rk_0'\right).
\label{eq:dMgeom}
\end{equation}
Together with the analytic decay of the nonminimal Wald surface terms, its far-zone convergence shows no resolved $O(\epsilon^2)$ mass coefficient,
\begin{equation}
\delta M_{\rm ADM/Wald}^{(2)}=0
\label{eq:dMzero}
\end{equation}
within the numerical asymptotic resolution.

\begin{figure}[t]
\includegraphics[width=\columnwidth]{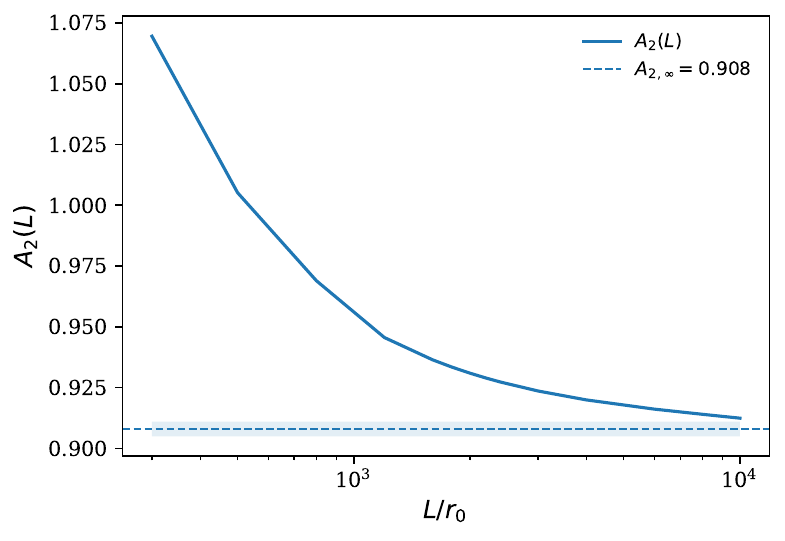}
\caption{Outer-domain convergence of $A_2(L)$, the fixed-$Q_\phi$ monopole coefficient defined by $A_{\rm th}/(4\pi r_0^2)=1+A_2\epsilon^2+O(\epsilon^4)$, as a function of the dimensionless outer cutoff $L/r_0$.}
\label{fig:area}
\end{figure}

At fixed $Q_\phi$, the throat area takes the form
\begin{align}
\frac{A_{\rm th}}{4\pi r_0^2}
&=1+A_2\epsilon^2+O(\epsilon^4),\\
A_2&=2k_0(0)\big|_{Q_\phi}=0.908\pm0.003.
\label{eq:areaeps}
\end{align}
The slow convergence of $A_2$ with the outer boundary is shown in Fig.~\ref{fig:area}.  In terms of the physical spin,
\begin{equation}
\frac{A_{\rm th}}{4\pi r_0^2}
=1+(7.74723\pm0.02560)\jhat^2+O(\jhat^4).
\label{eq:areaj}
\end{equation}

\section{Constrained quadrupolar sector}
\label{sec:l2}
With $P_2=P_2(\cos\theta)$, a compact form of the metric coframe through $O(\epsilon^2)$ is
\begin{align}
e^{\hat0}&=N\left[1+\epsilon^2(n_0+n_2P_2)\right]\dd t,\\
e^{\hat1}&=\left(1+\epsilon^2a_2P_2\right)\dd l,\\
e^{\hat2}&=R\left[1+\epsilon^2(k_0+k_2P_2)\right]\dd\theta,\\
e^{\hat3}&=R\left[1+\epsilon^2(k_0+k_2P_2)\right]\sin\theta\,\dd\phi
-\epsilon\frac{h}{R}\sin\theta\,\dd t .
\label{eq:full_coframe}
\end{align}
The scalar is
\begin{equation}
\phi=\phi_s+\epsilon^2\left(\phi_0+\phi_2P_2\right).
\end{equation}
In the same orthonormal frame the two-form is parameterized as
\begin{align}
B_{\hat0\hat1}&=-E_0-\epsilon^2\left(\delta E_{(0)}+\delta E_{(2)}P_2\right),\nonumber\\
B_{\hat1\hat3}&=\epsilon P_1,
& B_{\hat2\hat3}&=\epsilon Q_1,\nonumber\\
B_{\hat0\hat2}&=\epsilon^2 d_2\sin\theta\cos\theta,
\label{eq:full_B_perturbation}
\end{align}
where $P_1=\beta V\sin\theta/R$, $Q_1=\beta w\cos\theta/R^2$, and the fixed-norm corrections $\delta E_{(0,2)}$ are collected in Appendix~\ref{app:technical}.  Equation~\eqref{eq:full_B_perturbation} explicitly defines the polar two-form degree of freedom $d_2$.  In addition, a genuine Nambu--Goto bending mode appears because a rotated radial worldsheet is no longer described by fixed polar angle.  We parameterize the embedding by
\begin{equation}
\theta=\theta_0+\epsilon^2\chi_{\rm sc}(l)\partial_\theta P_2(\theta_0).
\label{eq:bending}
\end{equation}
Its action-derived bending equation is
\begin{align}
\frac{\dd}{\dd l}\left(\mu_{\rm eff}NR^2\chi_{\rm sc}'\right)
={}&\mu_{\rm eff}N(n_2+a_2)
+N\mu_{{\rm eff},\phi}\phi_2\nonumber\\
&+\frac{\mu_{\rm eff}h^2}{3NR^2}.
\label{eq:bendingeq}
\end{align}

The corrected second-order system is a constrained DAE.  Using accelerations $(a_2',n_2'',k_2'',d_2'',\phi_2'',\chi_{\rm sc}'')$, the six evolution rows $(E_{00},{\rm SC},{\rm BE},E_{22},E_{33},{\rm KR}_{02})$ have numerical principal rank five throughout the production domain.  The raw independent constraints are the radial Einstein equation $E_{11}$, the mixed $E_{l\theta}$ equation, and the angular-difference equation
\begin{equation}
G_2=E^\theta{}_\theta-E^\phi{}_\phi=0.
\end{equation}
The formal left-null relation among the six evolution rows is algebraically redundant with these three constraints together with the Bianchi and \KR{} identities; it is therefore not an additional boundary condition.  The average angular equation, rather than a magic-angle sample, supplies the correct scalar angular trace.

Reflection parity is imposed before interpolation.  In particular, $E_{11}$ and $G_2$ are even in the signed proper-radial coordinate and hence obey $E_{11}'(0)=G_2'(0)=0$.  Interpolating only on a half-domain can violate these derivative conditions and produces a spurious throat obstruction.  We therefore mirror all background and source functions with their exact parity and project each numerical transport step back onto the affine constraint manifold.  With this prescription the throat parity conditions close to machine precision; the final regular-throat/decaying-infinity subspaces intersect with full combined rank eight and a matching residual of approximately $1.65\times10^{-13}$.  Direct raw-equation substitution gives constraint residuals at the $10^{-14}$ level and evolution-equation residuals typically between $10^{-9}$ and $10^{-7}$ in absolute magnitude away from denominator-sensitive throat points.  Thus the apparent two-dimensional homogeneous kernel seen in unprojected or ill-conditioned transports is a numerical artifact rather than a physical degeneracy.

The final throat coefficient is
\begin{equation}
k_2(0)=-4.855020018059588.
\label{eq:k20}
\end{equation}
The invariant ratio of polar to equatorial throat circumference is
\begin{equation}
\frac{C_{\rm pol}}{C_{\rm eq}}
=1+\frac34 k_2(0)\epsilon^2+O(\epsilon^4),
\end{equation}
which becomes
\begin{equation}
\frac{C_{\rm pol}}{C_{\rm eq}}
=1-31.06797868\jhat^2+O(\jhat^4).
\label{eq:shapej}
\end{equation}
Thus slow rotation makes the throat oblate.

\section{Fractional Kalb--Ramond quadrupolar hair}
\label{sec:fractional}
Before analyzing the full matter-coupled tail, it is useful to identify the vacuum metric benchmark.  Setting $N=1$, $R=l$ and all localized matter perturbations to zero, an indicial ansatz $(n_2,a_2,k_2)\propto R^s$ gives the two vacuum powers $s_+=2$ and $s_-=-3$; the decaying eigenvector is $(n_2,a_2,k_2)\propto(-1,+1,+1)R^{-3}$.  The compact algebraic indicial relation is recorded in Appendix~\ref{app:technical}.  Finite-radius evaluation converges rapidly to these values, so any slower noninteger decay in the full solution must be a matter-induced asymptotic response rather than a shifted vacuum quadrupole exponent.

\begin{figure}[t]
\includegraphics[width=\columnwidth]{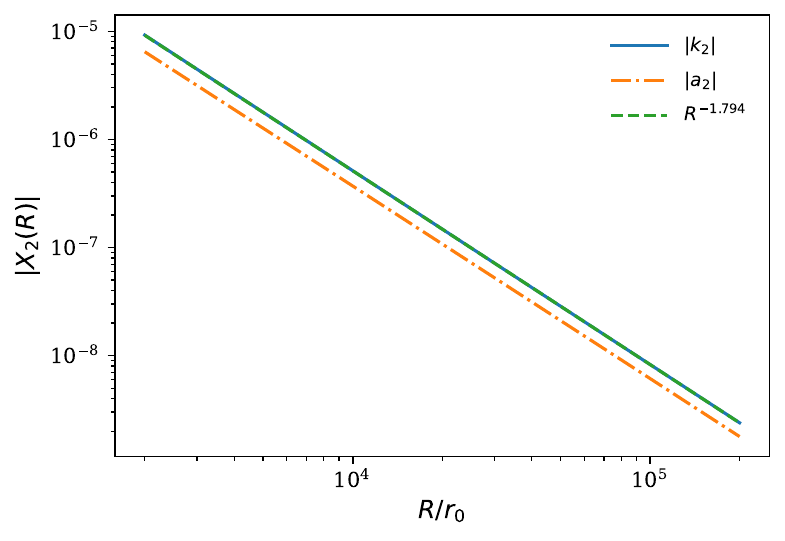}
\caption{Far-zone quadrupolar metric amplitudes.  The ordinate $|X_2|$ denotes either $|k_2|$ or $|a_2|$.  Both follow the noninteger power $R^{-q_{\rm KR}}$ with $q_{\rm KR}=1.793604973\ldots$, rather than the vacuum $R^{-3}$ quadrupole law.}
\label{fig:fractional}
\end{figure}

The far-zone structure differs qualitatively from a vacuum Hartle--Thorne hierarchy.  The localized static two-form decays as
\begin{equation}
\beta(R)=B_\infty R^{-p_+}[1+o(1)],
\qquad p_+=0.7936049731\ldots,
\end{equation}
whereas the regular polar response contains
\begin{equation}
d_2(R)=\frac{D_\infty}{R}+o(R^{-1}).
\label{eq:d2far}
\end{equation}
The purely first-order-squared Einstein sources decay as $R^{-6}$ and are subleading.  The leading effect instead comes from the nonminimal background--polar cross coupling.  Schematically,
\begin{equation}
\delta G_{\ell=2}[h_2]\sim\frac{H_2}{R^2},
\qquad
\delta T^{\rm KR,nm}_{\ell=2}\sim\frac{\beta d_2}{R^2},
\end{equation}
so that
\begin{equation}
H_2\sim\beta d_2\sim R^{-(1+p_+)}.
\label{eq:balance}
\end{equation}
We define
\begin{equation}
q_{\rm KR}=1+p_+=1.793604973109499.
\label{eq:qkr}
\end{equation}
The exponent in Eq.~\eqref{eq:qkr} is fixed by the benchmark localization root; other choices of $F(\phi)$ or $\mu_{\rm eff}(\phi)$ need not give the same numerical value.  The numerical far solution directly confirms
\begin{equation}
a_2,k_2\propto R^{-q_{\rm KR}},
\end{equation}
with clean-window amplitudes $k_2R^{q_{\rm KR}}\simeq-7.66$ and $a_2R^{q_{\rm KR}}\simeq5.6$.  Figure~\ref{fig:fractional} shows the measured slopes.  The physical perturbations are approximately
\begin{align}
\epsilon^2k_2
&\simeq-65.37\,\jhat^2
\left(\frac{r_0}{R}\right)^{q_{\rm KR}},\\
\epsilon^2a_2
&\simeq+47.97\,\jhat^2
\left(\frac{r_0}{R}\right)^{q_{\rm KR}}.
\label{eq:hairamplitudes}
\end{align}
The vacuum metric operator still contains the ordinary decaying solution
\begin{equation}
(n_2,a_2,k_2)\propto(-1,+1,+1)R^{-3},
\label{eq:vacq}
\end{equation}
but it is subleading to Eq.~\eqref{eq:balance}.  Therefore $R^3n_2$ or $R^3k_2$ does not approach an isolated quadrupole coefficient.

A curvature test confirms that the fractional term is not a pure gauge deformation.  At leading asymptotically flat order the $\ell=2$ harmonic of the linearized Ricci scalar is
\begin{align}
\delta{\cal R}_2={}&-2n_2''-4k_2''
+\frac{4a_2'-4n_2'-12k_2'}{R}\nonumber\\
&+\frac{12n_2+16a_2+8k_2}{R^2}.
\label{eq:ricci2}
\end{align}
Equation~\eqref{eq:ricci2} vanishes identically for Eq.~\eqref{eq:vacq}, but not for the fractional branch.  In the geometrically fixed asymptotic proper-distance/areal frame used in our construction we find
\begin{align}
r_0^2[\delta{\cal R}]_{\ell=2}
\simeq{}&(-191.1\pm2.6)\jhat^2
\left(\frac{r_0}{R}\right)^{3+p_+}
P_2(\cos\theta).
\label{eq:riccihair}
\end{align}
We use the coefficient in Eq.~\eqref{eq:riccihair} as a curvature-hair diagnostic in this asymptotic frame; the more robust statements are the nonzero fractional curvature tail and its exponent.

\section{Asymptotic charges and the multipole obstruction}
\label{sec:multipoles}
The fractional tail has an important consequence for conventional relativistic multipoles.  Let $\rho=1/R$ be a compactifying radial coordinate.  The leading quadrupolar spatial correction is
\begin{equation}
\rho^{q_{\rm KR}}P_2(\cos\theta),
\qquad 1<q_{\rm KR}<2.
\label{eq:compacttail}
\end{equation}
Near the compactified point, this is proportional in regular Cartesian coordinates to $\rho^{q_{\rm KR}-2}y^{\langle i}y^{j\rangle}$.  First derivatives are continuous, but generic second derivatives behave as
\begin{equation}
\rho^{q_{\rm KR}-2}=\rho^{-0.206395\ldots}.
\end{equation}
The compactified geometry is therefore generically $C^1$ but not $C^2$ at spatial infinity.

The standard Geroch--Hansen construction and its stationary non-vacuum extensions require sufficiently smooth conformal data at infinity \cite{Mayerson2023Multipoles}; similarly, the ACMC hierarchy reads mass multipoles from integer inverse powers $R^{-(\ell+1)}$ only when the corresponding smooth asymptotic expansion exists.  Equation~\eqref{eq:compacttail} precedes the $R^{-3}$ vacuum quadrupole and breaks that hierarchy.  We therefore do not assign a conventional isolated Geroch--Hansen/ACMC mass quadrupole to this branch.

This obstruction should not be confused with a failure of ADM asymptotic flatness.  The fractional metric tail decays faster than $1/R$, and both the ADM/Wald mass and angular momentum charges remain well defined.  What fails is the additional smoothness needed to organize the higher stationary field into the conventional integer-power multipole sequence.

\section{Domain of validity and limitations}
\label{sec:validity}
\begin{figure}[t]
\includegraphics[width=\columnwidth]{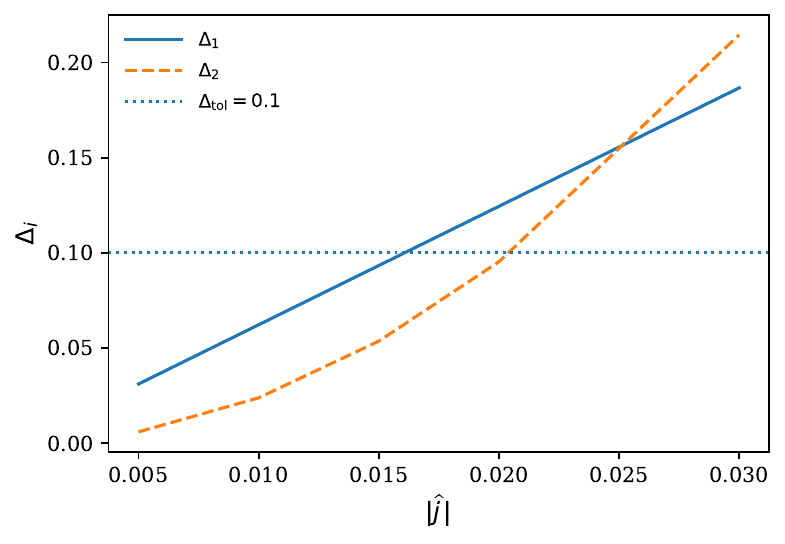}
\caption{Perturbative control as a function of $|\jhat|=|J|/r_0^2$.  We define $\Delta_1$ as the maximum first-order shift indicator and $\Delta_2\equiv\max|\epsilon^2 a_2|$ as the representative second-order correction.  The horizontal dotted line denotes the adopted truncation-control threshold $\Delta_{\rm tol}=0.1$, corresponding to a 10\% retained-correction criterion.}
\label{fig:validity}
\end{figure}

The construction is perturbative in $\epsilon$ and therefore in $\jhat$.  Some dimensionless second-order radial functions are numerically larger than unity: on the production branch, $\max|a_2|\simeq27.94$ and $\max|k_2|\simeq20.96$, while the largest first-order frame-dragging indicator is approximately $2.13$ before multiplication by $\epsilon$.  Figure~\ref{fig:validity} summarizes the retained corrections as a function of physical spin.  A conservative window for quantitative use is
\begin{equation}
|\jhat|\lesssim0.015\text{--}0.02.
\label{eq:validity}
\end{equation}
This is an accuracy criterion for the truncated series, not a dynamical stability bound.

\subsection{Numerical validation and frozen observables}
The numerical construction combines a high-accuracy static background, stable-manifold transport for the first-order axial system, and parity-aware projection-preserving transport for the second-order DAE.  The static production solution satisfies the representative residual levels $|E_{00}|\sim10^{-16}$, $|E_{11}|\lesssim6\times10^{-12}$ and $|E_{22}|\sim10^{-16}$, while the scalar Noether audit is at the $10^{-10}$ level.  At first order the outer stable-subspace matching residual is $1.3\times10^{-10}$, and at second order the projected matching residual is $1.65\times10^{-13}$.  Independent domain enlargement is used for the slowly convergent area and mass estimators.  Table~\ref{tab:results} collects the frozen observables used throughout the paper; quantities controlled by an outer-domain extrapolation are quoted with conservative numerical uncertainties.

\begin{table}[t]
\caption{Frozen benchmark observables used in the main text.}
\label{tab:results}
\begin{ruledtabular}
\begin{tabular}{lc}
observable & result\\
\hline
$M_{\rm ADM}^{(0)}$ & $0$\\
$\delta M_{\rm ADM/Wald}^{(2)}$ & $0$\\
$\delta\gamma^{(2)}$ & consistent with $0$\\
$J/(r_0^2\epsilon)$ & $0.3423494582$\\
$A_2$ & $0.908\pm0.003$\\
$k_2(0)$ & $-4.8550200181$\\
$C_{\rm pol}/C_{\rm eq}$ & $1-31.06797868\jhat^2+\cdots$\\
$q_{\rm KR}$ & $1.7936049731$\\
curvature-hair coefficient & $-191.1\pm2.6$\\
\end{tabular}
\end{ruledtabular}
\end{table}

Several limitations should be stated explicitly.  The regular stationary branch occurs at a tuned codimension-two coupling point and global uniqueness in $(\gamma,\xi_2)$ has not been proved.  The phantom scalar, the localization profile $F$, and the effective string tension in Eq.~\eqref{eq:mueff} are benchmark model ingredients rather than a UV-complete matter theory; the numerical value of $q_{\rm KR}$ is therefore model dependent even though the balance $q_{\rm KR}=1+p_+$ is structural for this branch.  Linear dynamical stability of the rotating branch remains open.  Rapid rotation requires a fully nonlinear continuation.  Finally, the absence of a conventional Geroch--Hansen quadrupole is an asymptotic consequence of the fractional tail, not a missing numerical extraction.

\section{Discussion}
\label{sec:discussion}
The present branch shares the Hartle--Thorne organization familiar from slowly rotating Ellis--Bronnikov wormholes, but its asymptotic response is qualitatively different \cite{SushkovKashargin2008,KasharginSushkov2008}.  In the phantom-scalar families, slow rotation leads to conventional monopole and quadrupole corrections and connects to fully nonlinear rotating solutions \cite{DzhunushalievEtAl2013,KleihausKunz2014,ChewKleihausKunz2016}.  Here the localized KR and string sectors leave the benchmark branch massless at the resolved orders while producing a long-range noninteger $\ell=2$ response.  The natural dimensionless spin is therefore $J/r_0^2$, not the black-hole-like ratio $J/M^2$.  Relative to earlier static KR wormholes and KR black holes, including recent deformed and matter-coupled KR geometries \cite{LessaSilvaMalufAlmeida2020,LessaOliveiraSilvaAlmeida2021,MalufMuniz2022,Yang2023,LiuWuWang2024,LiuWuWang2025,LiuWuWei2025,AraujoFilho2025NonCommutativeKR,AhmedSilva2026DyonicModMax}, the main new element is not simply the presence of rotation but the action-derived axial and polar response of the two-form together with the localized string sector.  Those matter responses are essential: omitting them would remove the channel that generates the fractional asymptotic tail.

The charge analysis provides an independent check of this interpretation.  Because the theory is diffeomorphism invariant but curvature coupled, the physical angular momentum is the Hamiltonian/Noether charge associated with the axial Killing vector \cite{LeeWald1990,Wald1993,IyerWald1994}.  For the regular branch, the nonminimal curvature contributions decay at infinity and the matter-sector surface terms do not contribute at $O(\epsilon)$, giving $J_{\rm Wald}=J_{\rm metric}$.  The same localization causes the static mass estimators and the resolved $O(\epsilon^2)$ mass correction to vanish asymptotically.  Thus the absence of an ADM mass is supported by both geometric estimators and the covariant surface-charge falloff, rather than by a single lapse fit.  This is important in nonminimally coupled systems, where metric parameters and covariant charges need not coincide automatically \cite{YuLyuHuheLi2026}.

The quadrupolar sector departs most strongly from the standard Hartle--Thorne picture.  The homogeneous vacuum operator contains the familiar decaying $R^{-3}$ solution, but the matter-supported response behaves as
\begin{equation}
 a_2,k_2\propto R^{-(1+p_+)},\qquad 1+p_+=1.7936049731\ldots .
\end{equation}
The source audit identifies the origin of this hierarchy: the quadratic axial Einstein sources decay as $R^{-6}$, while the background--polar KR product behaves as $\beta d_2\sim R^{-(1+p_+)}$ and therefore dominates the far-zone $\ell=2$ balance.  The same exponent survives in the Ricci scalar, so the effect is not a pure gauge deformation.  In inverted coordinates the compactified metric is generically $C^1$ but not $C^2$ at spatial infinity.  Since standard Geroch--Hansen/Thorne multipoles require sufficiently smooth asymptotic data, we characterize the leading rotational quadrupolar hair by the fractional exponent and the frame-qualified curvature-tail amplitude rather than by assigning a formal conventional quadrupole \cite{CompereOliveriSeraj2018,CanoGanchevMayerson2022,Mayerson2023Multipoles}.

The calculation also shows why action consistency matters at second order.  The $\ell=2$ equations form an index-one differential--algebraic system: five independent accelerations are accompanied by three constraint equations.  Evolution away from the constraint manifold generates spurious homogeneous freedom, whereas projection-preserving integration selects the regular boundary-value branch.  The string-bending equation plays the corresponding role in the material sector.  Because the string cloud arises from a Nambu--Goto congruence \cite{Letelier1979,SmalleyKrisch1997}, its worldsheets cannot consistently be held exactly radial once the quadrupolar geometry is perturbed.

The present analysis is not a stability theorem.  Static ghost-supported wormholes possess unstable modes \cite{ShinkaiHayward2002,GonzalezGuzmanSarbach2009}, while rotating studies indicate that angular momentum can weaken or remove such instabilities in parts of parameter space \cite{DzhunushalievEtAl2013,AzadEtAl2024}.  Testing that possibility here requires perturbing the complete KR--scalar--string system, including the constrained two-form sector.  Likewise, the quoted range $|\jhat|\lesssim0.015$--$0.02$ is a truncation-control criterion rather than a stability bound.  A nonlinear continuation should determine whether the fractional tail persists at larger spin and whether ergoregions or additional causal restrictions appear.  The numerical value of $p_+$ is model dependent because it depends on the localization profile and effective string tension; the more structural mechanism is $\beta\sim R^{-p_+}$ together with $d_2\sim R^{-1}$, which yields $h_{\ell=2}\sim R^{-(1+p_+)}$.

\section{Conclusions}
\label{sec:conclusions}
We have constructed a regular, action-consistent slow-rotation branch of a localized Kalb--Ramond wormhole through second order in the rotation parameter.  The stationary geometry, KR field, phantom scalar, and localized string sector are solved as one coupled system.  For the benchmark branch, the static ADM mass vanishes, no nonzero second-order ADM/Wald mass correction is resolved, and the nonminimal surface terms decay rapidly enough that the Iyer--Wald and metric angular momenta agree.  Rotation increases the throat area and makes the throat oblate within the conservative slow-rotation domain.

The main qualitative result is the nonstandard quadrupolar asymptotics.  The regular polar KR response couples to the localized background and produces a matter-supported noninteger tail that falls off more slowly than the ordinary vacuum $R^{-3}$ quadrupole.  The same hierarchy is present in curvature, reducing the smoothness of the conformal completion and preventing the standard smooth Geroch--Hansen/ACMC quadrupole construction for this branch.  The fractional decay exponent and the frame-qualified curvature-tail amplitude therefore provide the appropriate leading descriptors of its rotational quadrupolar hair, while the global mass and angular-momentum charges remain well defined.  Dynamical stability, robustness under alternative localization profiles, global uniqueness in the coupling plane, and continuation to rapid rotation remain open problems.

\begin{acknowledgments}
S.M. gratefully acknowledges support from Grant FL--10425067111 of the Agency of Innovative Developments of Uzbekistan.
\end{acknowledgments}

\appendix

\section{Technical formulae}
\label{app:technical}
For completeness, we collect here the longer perturbative formulae that are required by the action but would interrupt the main physical discussion.  Holding the metric fixed while varying the two-form gives
\begin{equation}
\delta(B_{\alpha\beta}B^{\alpha\beta})
=2B^{\mu\nu}\delta B_{\mu\nu},
\end{equation}
which fixes the $2\xi_1RB_{\mu\nu}$ term in Eq.~\eqref{eq:KR_eom}.  If $E_0$ is the static orthonormal electric amplitude and
\begin{equation}
P_1=\frac{\beta V}{R}\sin\theta,
\qquad
Q_1=\frac{\beta w}{R^2}\cos\theta,
\end{equation}
the $O(\epsilon^2)$ fixed-norm electric corrections are
\begin{align}
\delta E_{(0)}={}&\frac{b_0^2F_\phi\phi_0}{4E_0}
+\frac{\beta^2V^2}{3E_0R^2}
+\frac{\beta^2w^2}{6E_0R^4},\\
\delta E_{(2)}={}&\frac{b_0^2F_\phi\phi_2}{4E_0}
-\frac{\beta^2V^2}{3E_0R^2}
+\frac{\beta^2w^2}{3E_0R^4}.
\label{eq:appendix_fixednorm}
\end{align}
The quadratic first-order field-strength sources entering the monopole and quadrupole equations are
\begin{equation}
T_H^{(0)}=\frac{\bar Q^2}{6\xi_2R^4},
\qquad
T_H^{(2)}=\frac{\bar Q^2}{3\xi_2R^4}.
\label{eq:appendix_Hsources}
\end{equation}

For the vacuum far-zone diagnostic, substituting $(n_2,a_2,k_2)\propto R^s$ into the asymptotic metric subsystem gives
\begin{equation}
\det{\cal M}(s)\propto(s-2)(s+3),
\qquad
s_+=2,\quad s_-=-3,
\label{eq:appendix_indicial}
\end{equation}
with the decaying eigenvector
\begin{equation}
(n_2,a_2,k_2)\propto(-1,+1,+1)R^{-3}.
\label{eq:appendix_vacuum_mode}
\end{equation}
These equations provide the vacuum reference used in Sec.~\ref{sec:fractional}; the physical noninteger branch instead follows from the localized background--polar \KR{} coupling.

\bibliographystyle{apsrev4-2}
\bibliography{reference}

\end{document}